\documentclass[12pt,preprint,natbib,iop]{emulateapj} %Use this for all other purposes (emulateapj)
\usepackage{graphicx,epsfig,amsmath} %Comment out hyperref for submission

\lefthead{van der Wel et al.}
\righthead{High-Redshift Lens in CANDELS}
\slugcomment{Version: \today}
\begin{document}

\newcommand{\gala}{{\tt GALAPAGOS}~}
\newcommand{\gf}{{\tt GALFIT}~}
\newcommand{\mh}{H_{\rm{F160W}}}
\newcommand{\kms}{\>{\rm km}\,{\rm s}^{-1}}
\newcommand{\reff}{R_{\rm{eff}}}
\newcommand{\msol}{M_{\odot}}
\newcommand{\msola}{10^{11}~M_{\odot}}
\newcommand{\msolb}{10^{10}~M_{\odot}}
\newcommand{\msolc}{10^{9}~M_{\odot}}

\title{Discovery of a Quadruple Lens in CANDELS with a Record Lens
  Redshift $z=1.53$}

%\footnote{Based on observations
%    with the Hubble Space Telescope, obtained at the Space Telescope
%    Science Institute, which is operated by AURA, Inc., under NASA
%    contract NAS 5-26555; in part on observations collected at the
%    European Southern Observatory, Chile (168.A-0485); and in part on
%    observations made with the Spitzer Space Telescope, which is
%    operated by the Jet Propulsion Laboratory, California Institute of
%    Technology, under NASA contract 1407.}

\author{A.~van der Wel\altaffilmark{1}, 
G.~van de Ven\altaffilmark{1}, 
M.~Maseda\altaffilmark{1},
H.W.~Rix\altaffilmark{1}, 
G.H.~Rudnick\altaffilmark{2,1},
A.~Grazian\altaffilmark{3},
S.L.~Finkelstein\altaffilmark{4},
D.~C.~Koo\altaffilmark{5},
S.M.~Faber\altaffilmark{5},
H.C.~Ferguson\altaffilmark{6},
A.M.~Koekemoer\altaffilmark{6},
N.A.~Grogin\altaffilmark{6},
D.D.~Kocevski\altaffilmark{7}
}

\altaffiltext{1}{Max-Planck Institut f\"ur Astronomie, K\"onigstuhl
  17, D-69117, Heidelberg, Germany; e-mail:vdwel@mpia.de}

\altaffiltext{2}{Department of Physics and Astronomy, The University
  of Kansas, Malott room 1082, 1251 Wescoe Hall Drive, Lawrence, KS
  66045, USA}

\altaffiltext{3}{INAF - Osservatorio Astronomico di Roma, via Frascati
  33, 00040, Monteporzio, Italy}

\altaffiltext{4}{Department of Astronomy, The University of Texas at
  Austin, Austin, TX 78712, USA}

\altaffiltext{5}{UCO/Lick Observatory, Department of Astronomy and
  Astrophysics, University of California, Santa Cruz, CA 95064, USA}

\altaffiltext{6}{Space Telescope Science Institute, 3700 San Martin
  Drive, Baltimore, MD 21218, USA}

\altaffiltext{7}{Department of Physics and Astronomy, University of
  Kentucky, 505 Rose Street, Lexington, Kentucky 40506, USA}

\begin{abstract}
  Using spectroscopy from the \textit{Large Binocular Telescope} and
  imaging from the \textit{Hubble Space Telescope} we discovered the
  first strong galaxy lens at $z_{\rm{lens}}>1$.  The lens has a
  secure photometric redshift of $z=1.53\pm0.09$ and the source is
  spectroscopically confirmed at $z=3.417$.  The Einstein radius
  ($0.35$''; 3.0 kpc) encloses $7.6\times\msolb$, with an upper limit
  on the dark matter fraction of 60\%.  The highly magnified
  (40$\times$) source galaxy has a very small stellar mass ($\sim
  10^8~\msol$) and shows an extremely strong [OIII]$_{5007\rm{\AA}}$
  emission line ($EW_0\sim 1000\rm{\AA}$) bolstering the evidence that
  intense starbursts among very low-mass galaxies are common at high
  redshift.
\end{abstract}

\section{Introduction}\label{sec:intro}
Strongly lensing galaxies provide a range of important applications,
from the measurement of cosmological parameters through time delays
between lensed images \citep{refsdal64}, to the direct measurement of
total galaxy masses \citep[e.g.,][]{kochanek95}, to the evolution and
ages of elliptical galaxies \citep[e.g.,][]{rusin03, vandeven03}, to
the radial profile of the dark matter distribution
\citep[e.g.,][]{treu04}.

Since the discovery of the first $z=1$ strong lens almost 30 years ago
\citep{lawrence84,schneider86}, no strong lenses at higher redshifts
have been found, despite the large number of $z\lesssim1$ lenses
discovered since then \citep[e.g.,][]{bolton06, faure08, more12}, and
a handful of tentative $z\sim 1.2$ candidates \citep{more12} for which
the lens nature is unfortunately doubtful.

The paucity of $z>1$ lenses has four causes: first, lensing
probabilities decrease for lenses at higher redshifts due to the small
relative distances between lenses and sources -- for example, the
Einstein radius for a $\sigma = 200~\rm{km}~\rm{s}^{-1}$ galaxy at
$z=1.5$ is only $\sim0.4$'' for a $z>2.5$ source.  Second, the reduced
volume and number density of sources behind potential lenses is
greatly reduced.  Third, massive galaxies are increasingly rare at
higher redshifts, and those with the highest lensing probability --
the most concentrated, bulge-dominated galaxies -- are usually red and
faint in the rest-frame ultraviolet and hence in optical $z>1$
surveys.  Fourth, the measurement of lens and source redshifts and the
confirmation of the lens nature require near-infrared spectroscopy,
which has been much more challenging than optical spectroscopy.

Based on the number counts of potential high-redshift ($z>2.5$)
sources \citep[e.g.,][]{bouwens07} and the $z=1-2$ mass function of
potential lenses \citep[e.g.,][]{ilbert10,brammer11} we estimate one
$z>1$ galaxy-galaxy lens to occur per $\sim 200$ square
arcminutes. Thus, large-area, near-infrared surveys are needed to have
any chance to find $z>1$ lenses: the largest current survey of this
kind, CANDELS \citep{grogin11, koekemoer11}, covers 800 square
arcminutes with \textit{Hubble Space Telescope} near-infrared imaging,
such that it should contain a handful of $z>1$ lenses with background
sources intrinsically brighter than 28th magnitude (AB) in the
near-infrared.

Here we present the discovery of the quadruple galaxy-galaxy lens
J100018.47+022138.74 (J1000+0221) using imaging data from CANDELS and
near-infrared spectroscopy from the \textit{Large Binocular Telescope}
(LBT).  We report a spectroscopically confirmed source redshift of
$z_S=3.417\pm0.001$ and a record lens galaxy redshift of
$z_L=1.53\pm0.09$.

The $z=1.53$ lens is a flattened, quiescent galaxy with a stellar mass
of $\sim 6\times\msolb$ and appears to be located in a previously
unknown overdense region populated with at least a half a dozen of
$L^*$ galaxies at the same redshift.  Its Einstein radius of $0.35$''
(3.0 kpc at $z=1.53$) presents us with the first opportunity for a
direct mass measurement in this redshift range through lensing,
providing an important confirmation of previous results based on
stellar absorption line kinematics \citep[see, e.g.,][ and references
therein]{vandesande13}, which has been technically challenging.

%The $z=3.417$ source displays an extremely luminous
%[OIII]$_{5007\rm{\AA}}$ emission line.  It falls in the category of
%star-bursting dwarf galaxies, which have been shown to be abundant at
%$z\sim 2$ \citet{vanderwel11b}, and of which another specimen (at
%$z=1.8$) is strongly lensed by a $z=0.6$ early-type galaxy in the
%CANDELS UDS field \citep{brammer12b}.

%\begin{figure*}[t]
%\includegraphics[scale=.62,angle=90]{fig1v2f.ps}
%\caption{Quadrupole gravitational lens J1000+0221. The false-color
%  image is generated using HST/WFC3 imaging from CANDELS (F125W in
%  green, and F160W in red) and HST/ACS imaing from COSMOS and CANDELS
%  (F814W in blue).  The lensing galaxy is a lenticularly shaped galaxy
%  at $z=1.53\pm 0.09$ with little or no star formation and a stellar
%  mass of $\sim 5\times\msolb$.  The source is a young, strongly
%  starbursting galaxy with a very small stellar mass ($\sim
%  10^8~\msol$), magnified 40 times by the foreground galaxy, which is
%  the highest-redshift strong lens discovered to date. The zoomed-out
%  view shows the lens environment, which contains 6 $K_{AB}<22$
%  galaxies within 30 arcsec ($\sim 250$kpc) and photometric redshifts
%  $z\sim 1.55$.  One of these galaxies has a confirmed spectroscopic
%  redshift as indicated.  The most massive galaxy ($3-4\times \msola$)
%  is located slightly below the center of the image.}
%\label{fig:lens}
%\end{figure*}

\begin{figure}[h]
\epsscale{1} 
%\epsscale{0.5} 
%\plotone{14l_color.eps}
\plotone{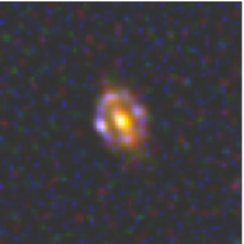}
\epsscale{2.1}
\plottwo{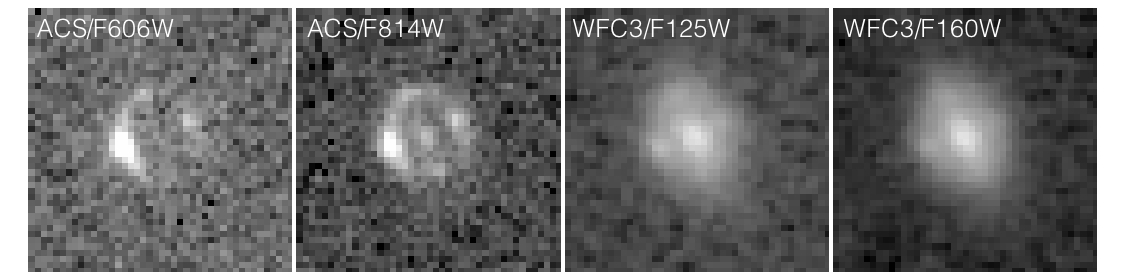}{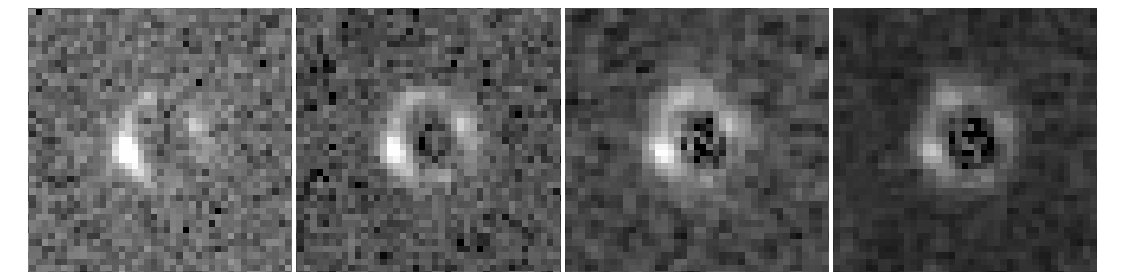}
\caption{\textit{Top}: Quadruple gravitational lens J1000+0221 with an
  Einstein radius of $0.35$''. The false-color image is generated
  using HST/WFC3 imaging from CANDELS (F125W in green, and F160W in
  red) and HST/ACS imaging from COSMOS and CANDELS (F814W in blue).
  These individual images were processed with the Lucy-Robertson
  deconvolution algorithm.  The lensing galaxy is a flattened galaxy
  with little or no star formation and a stellar mass of $\sim
  6\times\msolb$.  Its redshift of $z=1.53\pm 0.09$ makes this galaxy
  the most distant strong galaxy lens discovered to date. The source,
  magnified by a factor 40, is a young, strongly starbursting galaxy
  with a very small stellar mass ($\sim 10^8~\msol$). \textit{Bottom:}
  HST/ACS and HST/WFC3 images (2.4''$\times$2.4'') of the lens-source
  system (top row).  A S\'ersic model is produced based on the F160W
  image, masking the annulus containing the source images.  This
  S\'ersic model is scaled and subtracted from the other images.  The
  residuals are shown in the bottom row.}
\label{fig:lens}
\end{figure}

%\begin{figure}[h]
%\epsscale{2.4}
%\plottwo{fig2a.jpg}{fig2b.jpg}
%%\epsscale{0.5} 
%%\includegraphics[scale=.6,angle=-90]{fig2v2f.ps}
%\caption{  HST/ACS and HST/WFC3 images (2.4''$\times$2.4'') of the
%  lens-source system (top row).  A S\'ersic model is produced based on
%  the F160W image, masking the annulus containing the source images.
%  This S\'ersic model is scaled and subtracted from the other images.
%  The residuals are shown in the bottom row. Color variations among
%  the source images can be clearly seen.}
%\label{fig:imres}
%\end{figure}

\section{Data}\label{sec:data}
CANDELS \textit{HST}/Wide Field Camera 3 \citep[WFC3,][]{kimble08}
(F125W+F160W) and HST/Advanced Camera for Surveys
\citep[ACS,][]{ford03} (F606W+F814W) imaging of the lens+source system
is shown Figure \ref{fig:lens}.  The false-color image shows the
unambiguous lens nature of J1000+0221: the four blue images at a
sub-arcsecond scale surrounding a red galaxy are a tell-tale sign.
Before generating the color map each individual image was deconvolved
with the Lucy-Robertson algorithm, using appropriate point spread
functions (PSFs) for each filter (\S\ref{sec:decomp}).  Images in the
individual filters are shown in Figure \ref{fig:lens}. Especially
after subtraction of the lens (see \S\ref{sec:decomp} for details) the
almost continuous Einstein ring with a radius of $\sim 0.35$'' is
clearly seen in all filters.

\begin{figure}[t]
\epsscale{1.2} 
%\epsscale{0.5} 
\plotone{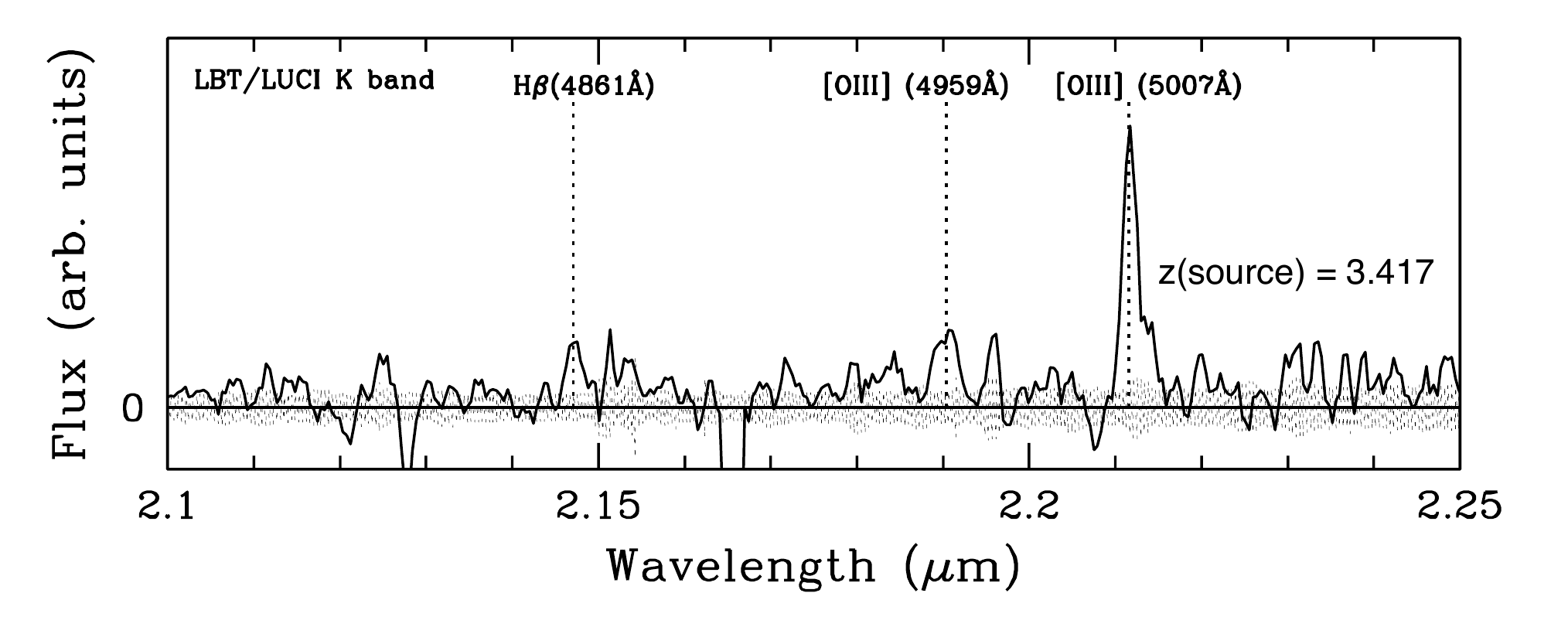}
\caption{LBT/LUCI K-band spectrum of J1000+0221, smoothed with the
  instrumental resolution.  The marginally non-zero continuum is
  mostly due to the $z=1.53$ lens. The three indicated nebular
  emission lines, all detected with 10$\sigma$ confidence or better,
  are from the source at $z=3.417$.  The strength of the emission line
  galaxies and [OIII]/H$\beta$ line ratio imply a very young (10 Myr),
  star-bursting system.  Given the magnification of $\mu=40$, we infer
  a stellar mass of $\sim 10^8~\msol$.}
\label{fig:sourcez}
\end{figure}

The source redshift $z_S=3.417$ is measured from LBT/LUCI
\citep{seifert03} near-infrared spectroscopy originally aimed at
obtaining continuum spectroscopy of massive galaxies at $z=1.5-2$.  A
3-hour observing sequence of individual, dithered 120s exposures with
the H+K grism in a 1" wide slit and seeing $\sim$0.6'' produced
significant detections of three emission lines in the K band,
identified as $H_{\beta}$ and $2\times$ [OIII] (see Figure
\ref{fig:sourcez}).  These fortuitously lensed emission lines allowed
us to identify the lens nature of the system, which we were able to
confirm upon visual inspection of the HST imaging described above.  No
spectroscopic redshift for the lens could be determined from the H- or
K-band spectrum.  The multi-slit capability of LUCI allowed us to
target several of the galaxies in the immediate vicinity of
J1000+0221, one of which we spectroscopically confirmed at $z=1.525$
through the detection H$\alpha$ and [NeIII] in the H band.  The LUCI
data reduction procedure is described in detail by M.~Maseda et
al.~(in prep.).

Finally, we include ground-based photometry in our analysis.  We use
the Newfirm Medium Band Survey (NMBS) multi-wavelength catalog and
derived data products from \citet{whitaker11}, as well as $Ugr$ LBT
photometry from the \textit{Large Binocular Cameras} (LBC, Boutsia et
al.~in prep.).  These datasets provide us with combined photometry for
the lens and source, which we will jointly analyze with the spatially
separated photometry from the HST data described in \S\ref{sec:decomp}
in order to obtain accurate redshift and stellar mass estimates.
%In addition, we
%use the photometric redshifts and stellar mass estimates of the
%surrounding galaxies to identify the over-density in which the lens is
%likely located.

\begin{center}
\begin{table*}
  \caption{Photometry based on decomposed HST images (Figure \ref{fig:lens}) and Spitzer/IRAC photometry.
  }
\begin{tabular}{l|cccccc|c}
\hline
\hline 
& F606W & F814W & F125W & F160W & IRAC$_{3.6}$ & IRAC$_{4.5}$ & z \\
& \multicolumn{6}{c|}{AB mag}    &   \\
\hline
lens   & 26.4$\pm$0.4 &  25.3$\pm$0.2 & 22.43$\pm$0.05 &  21.85$\pm$0.05 &  20.9$\pm$0.1 & 20.8$\pm$0.1 & 1.53$\pm$0.09   \\ 
source & 24.3$\pm$0.1 &  23.8$\pm$0.1 & 23.9$\pm$0.2 &  23.9$\pm$0.3 &  \nodata & \nodata & 3.417$\pm$0.001 \\
\hline
\end{tabular}
\label{tab:phot}
\end{table*}
\end{center}

\begin{table}\scriptsize
  \caption{Positions and flux ratios of lens and images. The positional uncertainty is $10^{-5}$ degrees.
  }
\begin{tabular}{l|ccc}
  \hline
  \hline 
  & R.A.  & Dec.  & F160W flux ratio \\
  & J2000 & J2000 & Arb. units \\
  \hline
  lens    & 150.0769694  & +2.3607623 & \nodata \\
  image A & 150.0770240  & +2.3608697 & 0.37$\pm$0.03 \\
  B       & 150.0768853  & +2.3607955 & 0.30$\pm$0.03 \\
  C       & 150.0770630  & +2.3607356 & 0.69$\pm$0.04 \\
  D       & 150.0769501  & +2.3606752 & 0.14$\pm$0.03 \\
  \hline
\end{tabular}
\label{tab:radec}
\end{table}

%\begin{table}\scriptsize
%  \caption{Lens model parameters: Einstein radius ($R_E$), mass enclosed in $R_E$ ($M_E$), isothermal velocity dispersion ($\sigma$), ellipticity of the mass distribution ($\epsilon$), total magnification of source ($\mu$).
%  }
%\begin{tabular}{l|c}
%  \hline
%  $R_E/\rm{kpc}$       &  3.0$\pm$0.2 \\
%  $M_E/\msolb$  & 7.6$\pm$0.5 \\
%  $\sigma/$(km s$^{-1}$)    & 182$\pm$10  \\
%  $\epsilon$  & 0.12$\pm$0.01  \\
%  $\mu$       & 40$\pm$2 \\
%  \hline
%\end{tabular}
%\label{tab:lens}
%\end{table}

%\begin{table}\scriptsize
%  \caption{Lens model parameters: Einstein radius ($R_E$), mass enclosed in $R_E$ ($M_E$), isothermal velocity dispersion ($\sigma$), ellipticity of the mass distribution ($\epsilon$), total magnification of source ($\mu$).  The last column list the adopted values based on avereging the results from Model 1 (image positions and flux ratios) and Model 2 (image positions only), and correcting for the contribution from large-scale structure for all parameters except the magnification (see text for details).
%  }
%\begin{tabular}{l|c|c|c}
%  \hline
%  \hline 
%  & Model 1 & Model 2 & Adopted \\
%  \hline
%  $R_E$       & 0.38''            &  0.35''             &  0.35''$\pm$0.02'' \\
%  $M_E/\msolb$       & $8.7$  & $7.6$  &  7.1$\pm$0.7 \\
%  $\sigma/$(km s$^{-1}$)    & 198    &  182    &  180$\pm$10 \\
%  $\epsilon$  & 0.12               & 0.12               &  0.13 $\pm$0.02 \\
%  $\mu$       & 37                 & 40                 &  39$\pm$2 \\
%  \hline
%\end{tabular}
%\label{tab:lens}
%\end{table}

\section{Lens-Source Image Decomposition}\label{sec:decomp}
We use \gf \citep{peng10} and the PSF from \citet{vanderwel12} to
produce a S\'ersic model for the F160W image of the lens.  An annulus
with an inner radius of 0.18'' and an outer radius of 0.66''
containing the source images is masked in the fitting process.  The
half-light radius measured along the major axis is $\reff = 0.29\pm
0.01$'' or $R_{\rm{eff}}=2.5\pm 0.1$~kpc.  The S\'ersic index is
$n=2.2\pm 0.2$, the projected axis ratio $q=0.52\pm 0.01$, and the
position angle ($PA=27\pm 1$~degrees from North to East).  The results
do not significantly change if the masked region is made narrower.
This size and shape are rather typical for early-type galaxies at
$z\sim 1.5$ \citep[e.g.,][]{vanderwel11a,chang13b}.

Then, for each of the ACS and WFC3 images, we convolve this S\'ersic
model with the appropriate PSF model\footnote{The F125W PSF is taken
  from \citet{vanderwel12}; a bright, non-saturated star in the mosaic
  serves as the PSF model for the ACS images.} and scale the total
flux to minimize the residual flux within a radius of 0.18''.  The
scaled models provide us with PSF-matched photometry of the lens.  The
lens-subtracted images are shown in the bottom row of Figure
\ref{fig:lens}.  Total source fluxes are then measured within 1''
diameter apertures.

In order to increase the wavelength coverage, we take the IRAC
3.6$\mu$m and 4.5$\mu$m photometry from the NMBS catalog and include a
small correction for the source flux, estimated to be 10\% based on
the spectral energy distribution of the source (see
\S\ref{sec:source}).  Hence we have obtained separate photometry for
the lens and the source, which we show in Figure \ref{fig:sed} and
list in Table \ref{tab:phot}.

\section{Lens Redshift and Stellar Mass}\label{sec:lens}

Spectroscopic confirmation of the lens redshift has remained elusive
due to its faintness ($V\sim 26$; $\mh\sim 22$) and apparent lack of
emission lines.  However, the red $I_{F814W}-J_{F125W}$ color implies
the presence of the Balmer and/or 4000$\rm{\AA}$ break straddled by
these two filters, suggesting a redshift in the range $1.2<z<2.0$.
Using {\tt EAZY} \citep{brammer08} we construct the redshift
probability distribution based on the photometry given in Table
\ref{tab:phot}, and find that the peak lies at $z=1.53$.  We find the
same result if we include the LBC $U$-band photometry.

\begin{figure}[t]
\epsscale{1.2} 
%\epsscale{0.5} 
\plotone{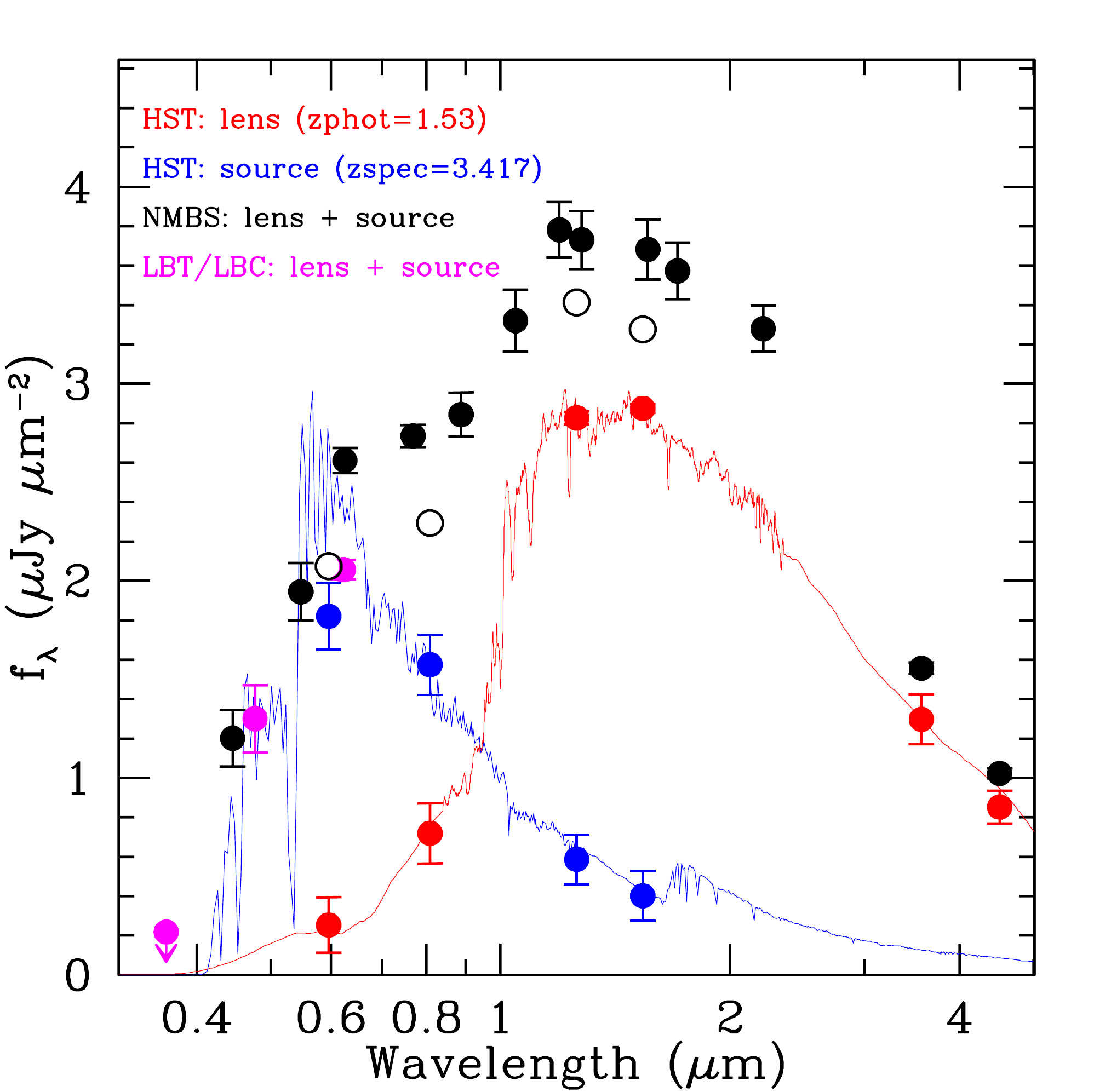}
%\plotone{zM.ps}
\caption{ Spectral energy distribution of gravitational lens system
  J1000+0221.  The blue points show the HST photometry of the source,
  derived from the lens-subtracted images shown in Figure
  \ref{fig:lens}; the red points show the HST photometry of the lens,
  obtained from \gf-based S\'ersic models for the lens; the open black
  points show the sum of the HST photometry for lens and source; the
  filled black points show ground-based photometry for the combined
  lens-source system from NMBS \citep{whitaker11}; the magenta points
  show ground-based photometry for the combined lens-source system
  from LBT/LBC (Boutsia et al. (in prep.).  The LBT/LBC fluxes
  $r$-band flux agrees very well with the HST F606W flux.  The offset
  between the NMBS photometry on the one hand, and the HST and LBT
  photometry on the other hand is of unknown origin.  The red and blue
  lines are spectral energy distributions that represent acceptable
  fits to the photometry of the lens and source, respectively.  The
  source model is a dust-free, 50 Myr old single stellar population
  (see \S\ref{sec:source}).  The lens model has exponentially
  declining star formation rate with an age of 1 Gyr and $A_V=0.8$
  (see \S\ref{sec:lens}).}
  %\textit{Right:} Stellar mass
  %vs.~redshift for all galaxies brighter than $K_{AB}=22$ and within
  %30\arcsec from the lens.  The thick (thin) error bars show the 68\%
  %(95\%) confidence intervals of the photometric redshift probability
  %distributions.  The lens itself is shown in red and the blue square
  %indicates the spectroscopic redshift for that galaxy.  Based on the
  %photometric redshift of the lens and the congregation of at least 6
  %massive galaxies with consistent photometric redshift, one of which
  %is spectroscopically confirmed at $z=1.525$, we conclude that the
  %lens is at $z=1.53\pm 0.09$ (68\% confidence).
\label{fig:sed}
\end{figure}

The lens-source system is unresolved in the ground-based imaging,
which thus produces the sum of the fluxes of the source and the lens.
The source has an accurately known redshift and a very simple spectral
energy distribution (SED), given that it is young and dust-poor as
indicated by the blue colors (see Figure \ref{fig:sed} and
\S\ref{sec:source}).  The {\tt EAZY} redshift probability distribution
for the lens based on the NMBS photometry after subtracting the source
SED peaks at $z=1.50$.  The combined probability distribution from the
two independent redshift estimates give $1.44<z<1.62$ ($1.34<z<1.74$)
as the 68\% (95\%) confidence interval, with no secondary redshift
solutions.

Further credence to our redshift estimate of the lens, and the precise
value of $z=1.53$ in particular, is lent by the four $\sim \msola$
galaxies with $z_{\rm{phot}}\sim 1.5$ in the NMBS catalog within a
projected distance of 250 kpc, one of which could be spectroscopically
confirmed at $z=1.525$ with our LUCI spectroscopy.  This is the only
such overdensity in the COSMOS field, and suggests that the lens is a
member.
% The contribution from large scale structure along the line of sight
% is a complicating factor in the interpretation of the lensing system
% (\S\ref{sec:lensmodel}).

We use {\tt FAST} \citep{kriek09} to estimate the lens stellar mass,
star-formation history and attenuation based on the ACS+WFC3+IRAC
photometry given in Table \ref{tab:phot}.  The best-fitting, solar
metallicity model, shown in red in Figure \ref{fig:sed}, has an age of
1 Gyr, with an exponentially declining star formation rate
($\tau=0.05$~Gyr), and moderate extinction (A$_{V}$=0.8).  Dustier
models with some residual star formation cannot be ruled out.  The
star formation rate is at most a few $\msol$ / yr$^{-1}$ given that we
see no H$\alpha$ in the LUCI spectrum and only have a marginally
significant Spitzer 24$\mu$m flux (from the NMBS catalog).  The
stellar mass is $M_* = 6^{+4}_{-1}\times \msolb$, assuming a
\citet{chabrier03} stellar initial mass function.  Although the
results of the SED fit suffer from the usual age-metallicity
degeneracy, the mass estimate is not sensitive to departures from the
adopted (solar) metallicity.

% In the regime where the light from the source dominates the total
% flux (at $<1\mu$m) we see $\sim20\%$ inconsistencies.  This is
% understandable, because the PSF-matched NMBS photometry uses the
% light distribution seen in the near-infrared to produce a model for
% the light distribution at other wavelengths, an approach that fails
% in the presence of strong morphological changes with wavelength.
% The K band is another outlier: here the luminous [OIII] emission
% line from the source significantly contributes to the total flux
% (Figure \ref{fig:sourcez}), which we will further explore in
% \S\ref{sec:source}.

\section{Lens Model}\label{sec:lensmodel}

We use the method described by \citet{vandeven10} to produce an
analytical lens model that allows for a non-spherical underlying mass
distribution.  In Table \ref{tab:radec} we provide the coordinates of
the lens center and the four source images, as well as flux ratios for
the images measured in F160W.  The flux ratios vary with wavelength,
as is immediately apparent from the the lens-subtracted images shown
in Figure \ref{fig:lens}: the brightest image is also the bluest.  We
rule out a supernova by examining the ACS images at two different
epochs, such that the color variations suggest that the source has
intrinsic, spatial color variations, or that there is dust in the
lens, the distribution of which is patchy.  Because of the color
variations we do not use the flux ratios of the images in our lensing
model.  Our best-fitting lens model has an Einstein radius of
$R_E=0.35$'' (or 3.0 kpc) with an enclosed mass of $M_E =
(7.6\pm0.5)\times\msolb$.  The total magnification of the source,
which is aligned almost perfectly (to within 0.01'') with the lens, is
$\mu=40\pm2$.
% The orientation of the mass distribution is consistent with the
% orientation of the light distribution, such that at least
% qualitatively mass follows light.

Given that 66\% of the F160W light from the lens falls within the
Einstein radius we derive a stellar mass within the Einstein radius of
$M_*(<R_E) = 4^{+2}_{-1}\times\msolb$, which implies that stars likely
account for the majority, but not necessarily all, of the mass within
$R_E$.  This is also the case if we increase the stellar mass by
adopting the Salpeter IMF (which cannot be ruled out) rather than the
Chabrier IMF.  An indication that light does not precisely trace mass
is that the (tangential) critical curve for a mass distribution that
is as flattened as the observed light distribution should also be
somewhat elongated (in the direction perpendicular to that of the
light distribution).  The image configuration, however, implies that
critical curve is very close to circulur.  Furthermore, if mass
follows light, then the implied mass profile at $R_E$ is unusually
steep: with little scatter, the typical slope is known to be close to
isothermal \citep[$\gamma \sim 2$,][]{koopmans09}, whereas the implied
slope here is $\gamma\sim 3$.  Given our stellar mass estimate, the
roundness of the critical curve, and the extremely steep mass profile
slope inferred under the assumption that mass follows light, we
conclude that there likely is a significant contribution from dark
matter within $R_E$ of the lens, with an 1-$\sigma$ upper limit of
60\%.

As described above, the lens is situated in an overdense environment,
which likely contributes to the projected, enclosed mass.  A crude
estimate of this effect can be inferred by assuming that the most
massive galaxy in the overdensity, which lies at a projected distance
of 110 kpc, is situated at the center of a spherical \citet{navarro96}
dark matter halo with concentration parameter $c=6$.  Based on the
stellar mass of this central galaxy ($M_{*} = 3\times\msola$) we
estiamte the halo to have a total mass of $10^{13.5-14}\msol$
\citep{moster13}, such that the projected mass within the Einstein
radius of the lens is $\sim\msolb$.  This does not significantly
affect our conclusions regarding the dark matter and stellar mass
fractions.

Finally, we note that the velocity dispersion implied by the lens
model ($\sqrt{M_E/R_E} \equiv \sigma_E=182\pm10$~km s$^{-1}$) is
consistent with stellar velocity dispersions of galaxies at similar
redshifts as measured from absorption line spectra \citep[e.g.,][and
references therein]{vandesande13}: the velocity dispersion is higher
by $\sim 20\%$ compared to present-day galaxies with a similar mass.
Our accurate and precise measurement of the velocity dispersion
provides important support to the results based on stellar kinematics
given the practical difficulties associated with continuum
spectroscopy for $z\gtrsim 1.5$ galaxies.

\section{Nature of the Source}\label{sec:source}
The HST photometry of the source (blue points in Figure \ref{fig:sed})
shows a very blue continuum, indicative of a young, dust-free galaxy.
Using the method described by \citet{finkelstein12} we fit the
spectral energy distribution given by the four photometric data points
(F606W, F814W, F125W, and F160W) given in Table \ref{tab:phot}.  We
confirm the young age ($50\pm30$~Myr), dust-free nature
($E(B-V)=0.00^{+0.16}_{-0.00}$), and blue continuum slope
($\beta=-2.0\pm0.3$) of the source.  The (de-magnified) stellar mass
of the galaxy is $M_*=2.6^{+2.0}_{-1.3}\times 10^8~\msol$.  The
best-fitting model spectral energy distribution is shown in Figure
\ref{fig:sed}.

As we mentioned above and showed in Figure \ref{fig:sourcez}, three
emission lines are detected with high significance: H$\beta$ and
[OIII] at 4959$\rm{\AA}$ and 5007$\rm{\AA}$.  We detect a low-level
(1-2$\sigma$ per pixel) continuum in the spectrum, and we use the
average continuum flux over the entire K-band wavelength range to
estimate the equivalent width of the brightest [OIII] line at
$EW([OIII]_{5007\rm{\AA}})=340\pm 50\rm{\AA}$ in the observed frame,
or $EW_0([OIII]_{5007\rm{\AA}})=77\pm 11\AA$ in the rest-frame of the
source\footnote{The lens+source system is smaller than the slit width,
  as is the seeing disk for our observations, such that slit losses
  are negligible.}.  Adopting the K band flux ratio predicted by the
SED models for the source and the lens we arrive at an estimated
rest-frame $EW_0([OIII]_{5007\rm{\AA}})=1200\pm 300\rm{\AA}$ for the
source alone.

This high $EW$ is consistent with the stellar population properties
estimates above, although we note that this is not necessarily the
$EW$ of the unlensed source, as a spatially varying $EW$ could lead to
a boosted, lensed $EW$.  The fact that the brightest image is also the
bluest may indicate that this may be the case here, but it is beyond
the scope of our analysis to attempt to correct for this effect.

The [OIII]$/$H$\beta$ line ratio from the spectrum shown in Figure
\ref{fig:sourcez} is $5^{+2}_{-1}$, indicative of a high excitation,
and likely a low metallicity.  Additional emission lines such as [OII]
or [NeIII], needed to confirm the low metallicity, are not detected.

An abundant population of objects at $z\sim 2$ with very high [OIII]
$EW_0$ was identified by \citet{vanderwel11b}, and the properties of
those extreme emission line galaxies are very similar to those of the
source described here \citep[also see][]{atek11}.  Remarkably, this is
the second such emission-line dominated object that is found to be
strongly lensed in CANDELS: using data from the 3D-HST survey
\citep{brammer12a}, \citet{brammer12b} found a galaxy at
$z_{\rm{S}}=1.847$ with very similar mass and age.  Given the
abundance of unlensed emission-line galaxies it was seemingly unlikely
that even one strongly lensed version exists within a CANDELS-sized
survey.  The existence of two strong lensed galaxies of this kind
suggests that such galaxies are likely even more common than is
currently assumed, and that objects near or below the detection limits
of current surveys -- the unlensed magnitude of the $z=3.417$ source
is $\mh\sim 28$ -- very commonly display such strong emission lines
\citep[\textit{cf.},][]{smit13}.  This would have profound
consequences for our understanding of galaxy formation: a presumably
brief burst of intense star formation may signal the initial formation
stage of any galaxy.

Euclid will reveal tens of thousands $z>1$ lens candidates, which will
give us unprecendented insight into the variety in properties among
faint, high-redshift galaxies, as well as the galaxy mass function at
all redshifts $0<z<2$ with perfectly calibrated mass measurements in
an absolute sense.

\acknowledgements{The authors thank Ricardo Amorin, Marcello Cacciato,
  Nimish Hathi, David Koo, Ray Lucas, Sharon Meidt, Mara Salvato,
  Benjamin Weiner, and Steven Willner for helpful comments and
  suggestions.  GHR acknowledges the support of an Alexander von
  Humboldt Foundation Fellowship for Experienced Researchers.}

\bibliographystyle{apj}

\end{document}